\documentclass[amssymb,amsmath,superscriptaddress,showpacs,floatfix]{revtex4-1}

\usepackage[dvips]{graphicx}
\usepackage{mathrsfs}
\usepackage{amsmath,amssymb}
\usepackage{hhline}
\usepackage{dcolumn}
\usepackage{bm}

\usepackage[dvips]{color}

\newcommand{\tphi}{\tilde{\phi}}
\newcommand{\kj}{{k^{\prime}}}
\newcommand{\kl}{{k^{\prime\prime}}}
\newcommand{\rD}{{\rm D}}
\newcommand{\rnD}{{\rm nD}}

\begin{document}

\title{Observability transitions in correlated networks}
\date{\today}
\author{Takehisa Hasegawa}
\affiliation{Graduate School of Information Science, Tohoku University,\\ 6-3-09, Aramaki-Aza-Aoba, Sendai, Miyagi, 980-8579, Japan}
\author{Taro Takaguchi}
\affiliation{National Institute of Informatics,\\ 2-1-2 Hitotsubashi, Chiyoda-ku, Tokyo, 101-8430, Japan}
\affiliation{JST, ERATO, Kawarabayashi Large Graph Project, Japan}
\author{Naoki Masuda}
\affiliation{Department of Mathematical Informatics, The University of Tokyo,\\ 7-3-1 Hongo, Bunkyo-ku, Tokyo, 113-8656, Japan}

\begin{abstract}
Yang, Wang, and Motter [\textit{Phys. Rev. Lett.} {\bf 109}, 258701 (2012)] analyzed a model for network observability transitions in which a sensor placed on a node makes the node and the adjacent nodes observable. The size of the connected components comprising the observable nodes is a major concern of the model. We analyze this model in random heterogeneous networks with degree correlation. With numerical simulations and analytical arguments based on generating functions, we find that negative degree correlation makes networks more observable. This result holds true both when the sensors are placed on nodes one by one in a random order and when hubs preferentially receive the sensors. Finally, we
numerically optimize networks with a fixed degree sequence with respect to the size of the largest observable component. Optimized networks have negative degree correlation induced by the resulting hub-repulsive structure; the largest hubs are rarely connected to each other, in contrast to the rich-club phenomenon of networks.
\end{abstract}

\pacs{89.75.Fb, 89.75.Hc, 64.60.aq}


\maketitle

\section{Introduction}

In power-grid networks, the state of a node, i.e., the complex voltage, can be determined by so-called phase measurement units
(PMUs; we call them \textit{sensors} in the following).
In a simplified setting, the state of a node is observed if a
sensor is placed on the node or its neighbor. In other words, a sensor is capable of
measuring the states of a node and its neighborhood. Because measurement is considered to be costly,
one is interested in completely or partly observing the nodes in a given
network with a small number of sensors~\cite{Johnson1974,Raz1997}.
Similar situations may occur in
dissemination of information or behavior in social networks. In this case,
the sensors correspond to information sources, for example, and a source node may be capable of affecting some neighboring nodes.

Recently, Yang and colleagues formulated this problem by employing the framework of
observability transitions~\cite{Yang2012}.
They determined the transition point with
respect to the density of sensors
above which the largest observable
component (LOC) is of a macroscopic size. The results were theoretically
derived in the
case of random placement of sensors and uncorrelated heterogeneous networks.
They also proposed a heuristic algorithm to determine the order in which
the nodes receive the sensors for
efficient observation of networks with a community structure.

Many networks exhibit degree correlation; the degrees of adjacent nodes
are not independent of each other~\cite{Newman2002,Serrano2007}.
The effect of degree correlation has been investigated for
various static and dynamic phenomenological models on networks. Examples include percolation~\cite{Newman2002,Vazquez2003,Noh2007,Goltsev2008,Shiraki2010,Herrmann2011,Schneider2011,Tanizawa2012},
susceptible-infected-susceptible~\cite{Eguiluz2002,Boguna2003,Boguna2003a} and susceptible-infected-recovered~\cite{Boguna2003a,Moreno2003,Hasegawa2012} models for epidemic spreading, synchronization~\cite{Wang2007,DiBernardo2007}, evolutionary game dynamics~\cite{Rong2007,Pusch2008}, and network controllability \cite{Posfai2013}. In particular, degree correlation is known to affect the robustness of networks against random failure of nodes or links. In many network models with degree correlation, negative degree correlation makes the critical node or link occupation probability above which a giant component emerges large, which makes the network less robust
\cite{Newman2002,Vazquez2003,Noh2007,Shiraki2010}.

Motivated by these previous studies,
in the present paper we examine the effect of degree correlation on
observability transitions.
We study the model of 
network observability transitions proposed in Ref.~\cite{Yang2012} in the
case of random and degree-based (i.e., hub-first) placement of sensors in
uncorrelated and correlated networks.
We have good reason to believe that degree correlation is related to observability transitions.
To enlarge the LOC for better observability, it is
intuitively clear that it is better to avoid allocating sensors to
adjacent nodes because such an allocation would yield redundant
observation of nodes by multiple sensors. At the same time,
nodes in heterogeneous networks
are likely to be efficiently observed if we preferentially put sensors on
hubs, similar to the role that hubs play in percolation
transitions (see Refs.~\cite{Dorogovtsev2008,Newman2010,Cohen2010} for reviews).
%
%
By combining these
two lines of consideration, we postulate that networks with negative
degree correlation such that hubs are separated from each other
enable efficient observation when the hubs preferentially
receive sensors. Even when sensors are sequentially placed on randomly selected nodes, the results may depend on the degree correlation of the network for the same reason.

\section{Model}

For different
placement rules and different networks,
we study the model introduced in Ref.~\cite{Yang2012}, which is defined as follows.
For a given network, we specify a subset of the nodes that are
directly observable by a sensor.
We denote the fraction of the directly observable nodes by $\phi$
($0\le \phi\le 1$).
Any node adjacent to at least one directly observable node is
also observable. We say that such a node is indirectly observable when it is not directly observable.
Then, a node is either directly observable, indirectly observable, or
unobservable. 
The problem of finding the smallest set of the directly observable
nodes that will make the entire network observable is known as the minimum dominating set problem in graph theory~\cite{Johnson1974,Raz1997} (also analyzed
in Ref.~\cite{Yang2012}).
Here, similar to the percolation problem, we focus on the size of the LOC,
defined as the largest connected component composed of directly or
indirectly observable nodes.

\section{Numerical simulations for networks with degree correlation}

In this section, we numerically investigate the relationships between
the degree correlation of networks and the size of the LOC.
We use scale-free networks, i.e., those with power-law degree
distributions, and Poisson networks, i.e., those with the Poisson degree distribution.
It should be noted that real power grids do not necessarily have power-law degree distributions~\cite{Amaral2000,Pagani2013}. However, we use scale-free networks to illustrate the effect of generally heterogeneous degree distributions, not to realistically model power grids. In addition, the World Wide Web and social networks, to which observability transitions are expected to be relevant, are scale-free (e.g., Ref.~\cite{Newman2010}).

\subsection{Generation of networks with a specified degree correlation}\label{sub:generation corr nets} 

We generate scale-free and Poisson networks with a specified degree correlation as follows.
First, we generate an uncorrelated network.
In the case of scale-free networks, we do so by using a growing
network model~\cite{Krapivsky2000,Dorogovtsev2000,Krapivsky2001a,Krapivsky2001b}. 
We start with two mutually
connected
nodes and add nodes one by one until the network has $N$
nodes. Each new node joins the existing
network with $m=2$ links. The
destinations of the new links are selected from the existing nodes
with the probability proportional to $k_i+k_0$, where $k_i$ is the
degree of the $i$th existing node and $k_0$ is a constant.  In this
way, we obtain a scale-free network with $p(k)\propto k^{-(3+k_0/m)}$
and mean degree $\left<k\right> \approx 2m = 4$. 
Then, we rebuild the network by using
the configuration model~\cite{Newman2010}, with which
the links are randomly rewired under the condition that
the degree of each node is conserved.
In the case of Poisson networks, we use the
Erd\H{o}s-Reny\'{i} (ER) random graph on $N$ nodes with 
$\left<k\right>\approx 4$ as the initial network. We connect each pair
of nodes with probability $4/(N-1)$, independently for different pairs.

To introduce degree correlation, we apply to the uncorrelated
scale-free or Poisson network a link swapping
algorithm~\cite{Maslov2002,Xulvi-Brunet2005,Newman2010}.
Specifically, we randomly select two links ($v_1$, $v_2$) and ($v_3$,
$v_4$) from the current network, where $v_i$ ($1\le i\le 4$) is a
node. Then, we tentatively replace links ($v_1$, $v_2$) and ($v_3$, $v_4$) by
($v_1$, $v_3$) and ($v_2$, $v_4$). We actually rewire the links if and
only if multiple edges and self-loops do not appear after the
rewiring and the new network has a more desirable degree correlation than
the old network does. Otherwise, we discard the proposed rewiring.
In either case, we repeat the rewiring attempts until a network with a
targeted degree correlation is realized.
It should be
noted that the rewiring preserves the degree of each node.

To decide whether to accept a proposed rewiring, we need to quantify
the degree correlation.
We measure the degree correlation using the Pearson correlation of the
excess degree~\cite{Newman2002}:
\begin{equation}
r = \frac{M^{-1} \sum_{i=1}^M K_i K_i^{\prime} - \left[ M^{-1}
    \sum_{i=1}^M \frac{1}{2} \left( K_i + K_i^{\prime} \right)
  \right]^2}{M^{-1} \sum_{i=1}^M \frac{1}{2} \left( K_i^2 +
    K_i^{\prime 2} \right)  - \left[ M^{-1} \sum_{i=1}^M \frac{1}{2} \left( K_i + K_i^{\prime} \right)\right]^2},
\end{equation}
where $K_i$ and $K_i^{\prime}$ are the degrees of the endpoints of the $i$th link ($1 \leq i \leq M$), and $M$ is the number of links in the network. It should be noted that $-1\le r\le 1$.

\subsection{Numerical results}\label{sub:numerical results}

We carry out numerical simulations with scale-free and Poisson networks with different $r$ values.
For scale-free networks, we set $p(k)\propto k^{-2.5}$ and
generate networks with $r=-0.1$, 0, and 0.1. The absolute
values of $r$ for correlated
networks are relatively small because it is logically difficult to
generate correlated heterogeneous networks with large $|r|$ values;
$r$ tends to 0 as $N\to\infty$~\cite{Serrano2007,Alderson2007,Holme2007PhysicaA,Menche2010,Dorogovtsev2010}.

\begin{figure}[!h]
 \begin{center}
  \includegraphics[width=75mm]{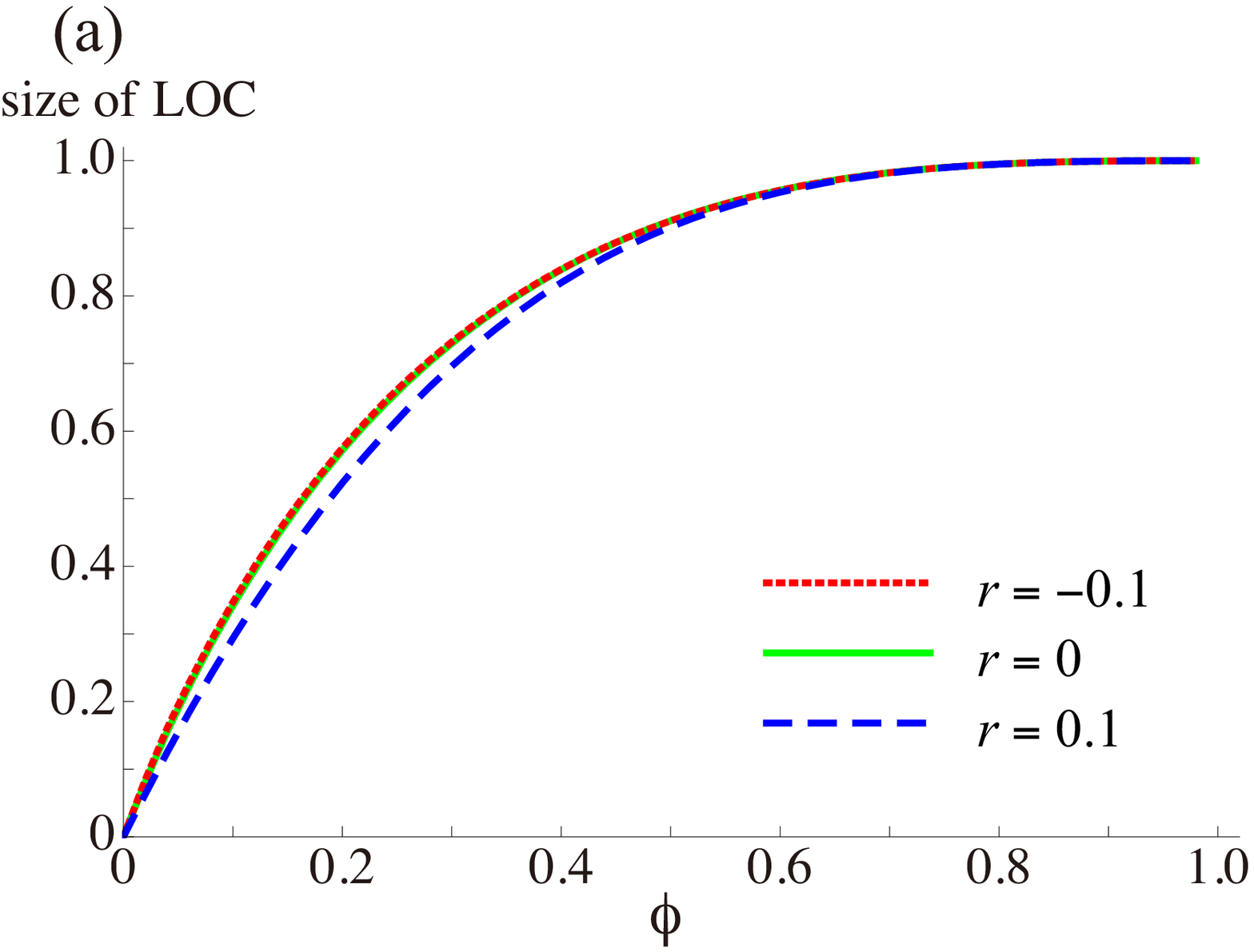}
  \includegraphics[width=75mm]{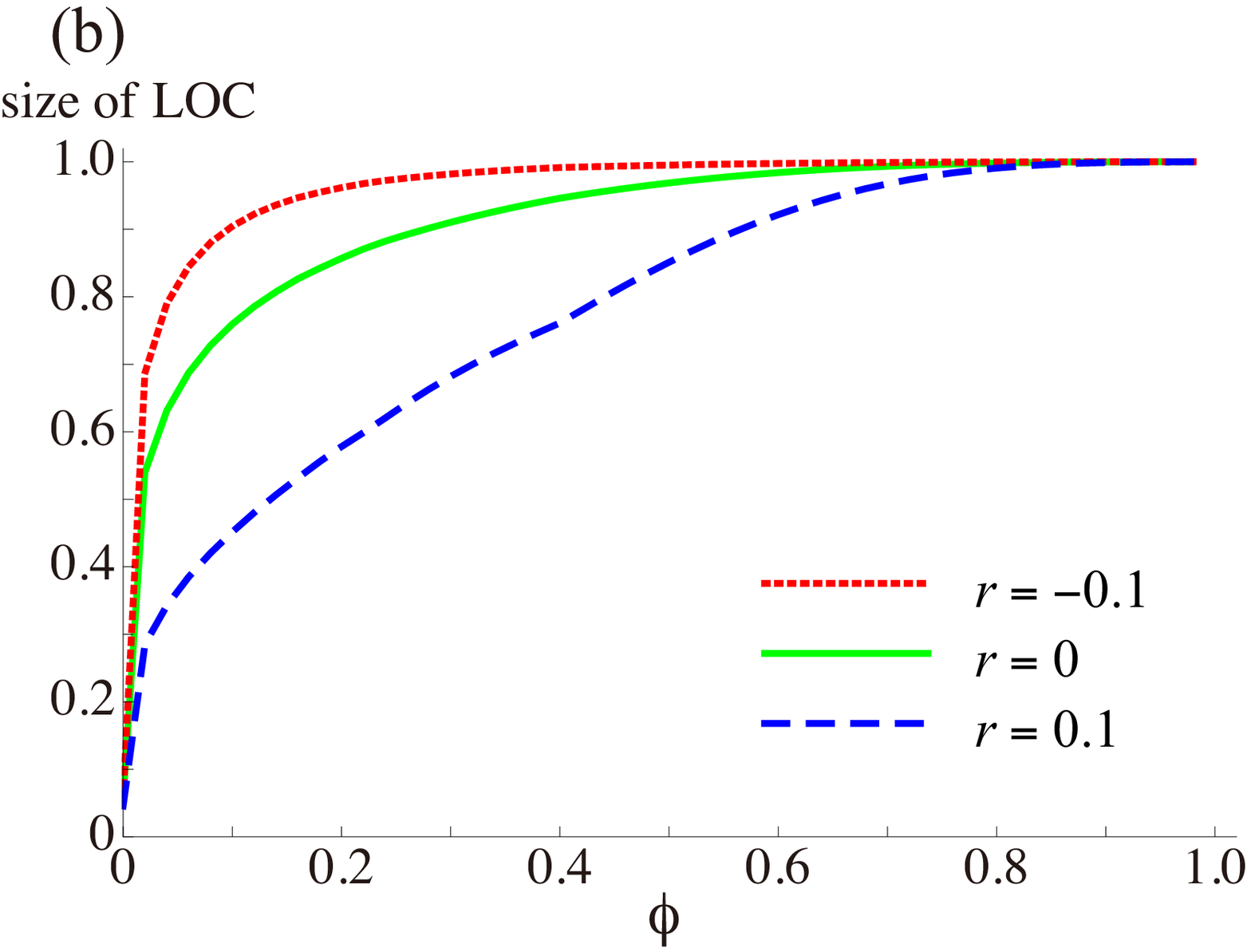}
  \includegraphics[width=75mm]{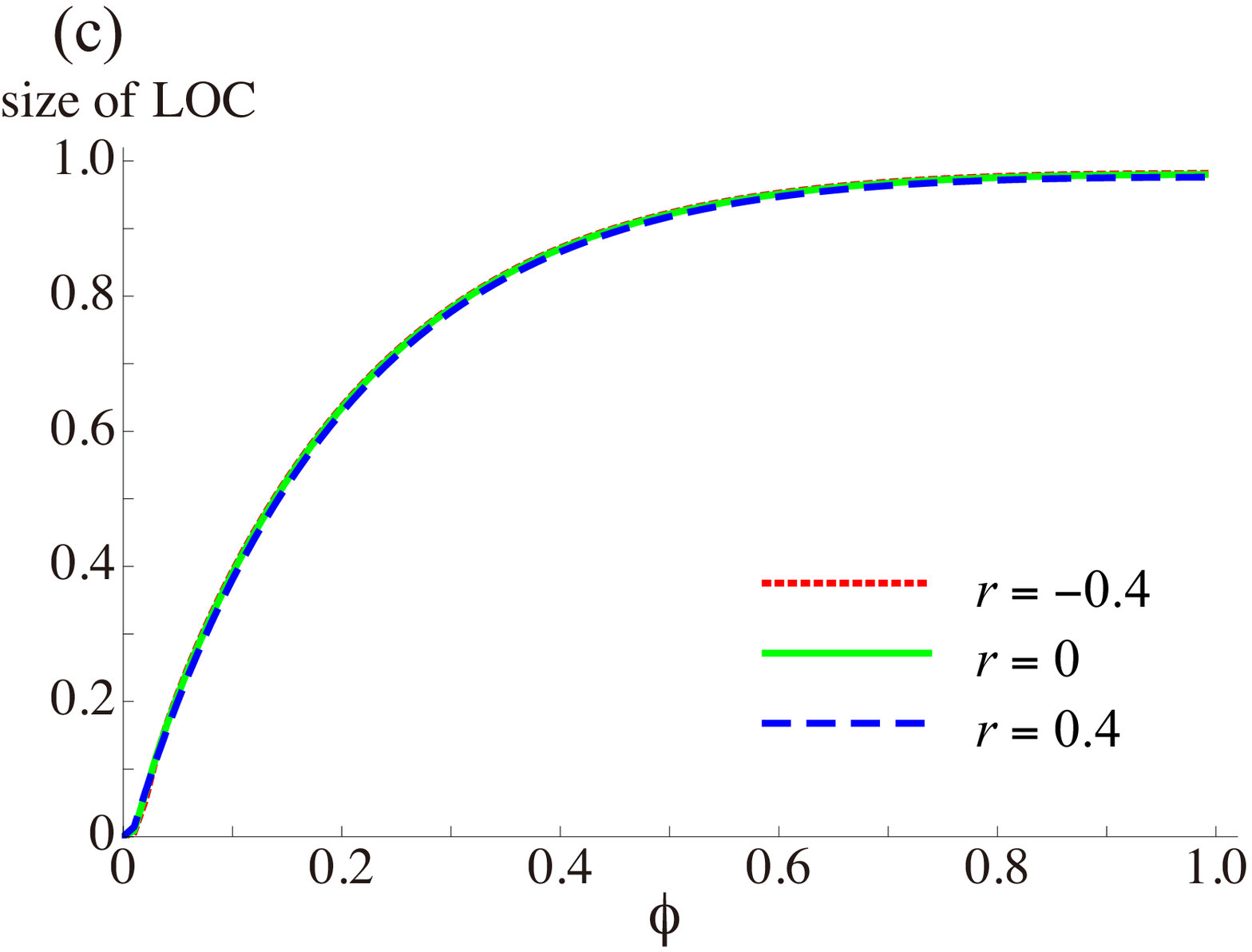}
  \includegraphics[width=75mm]{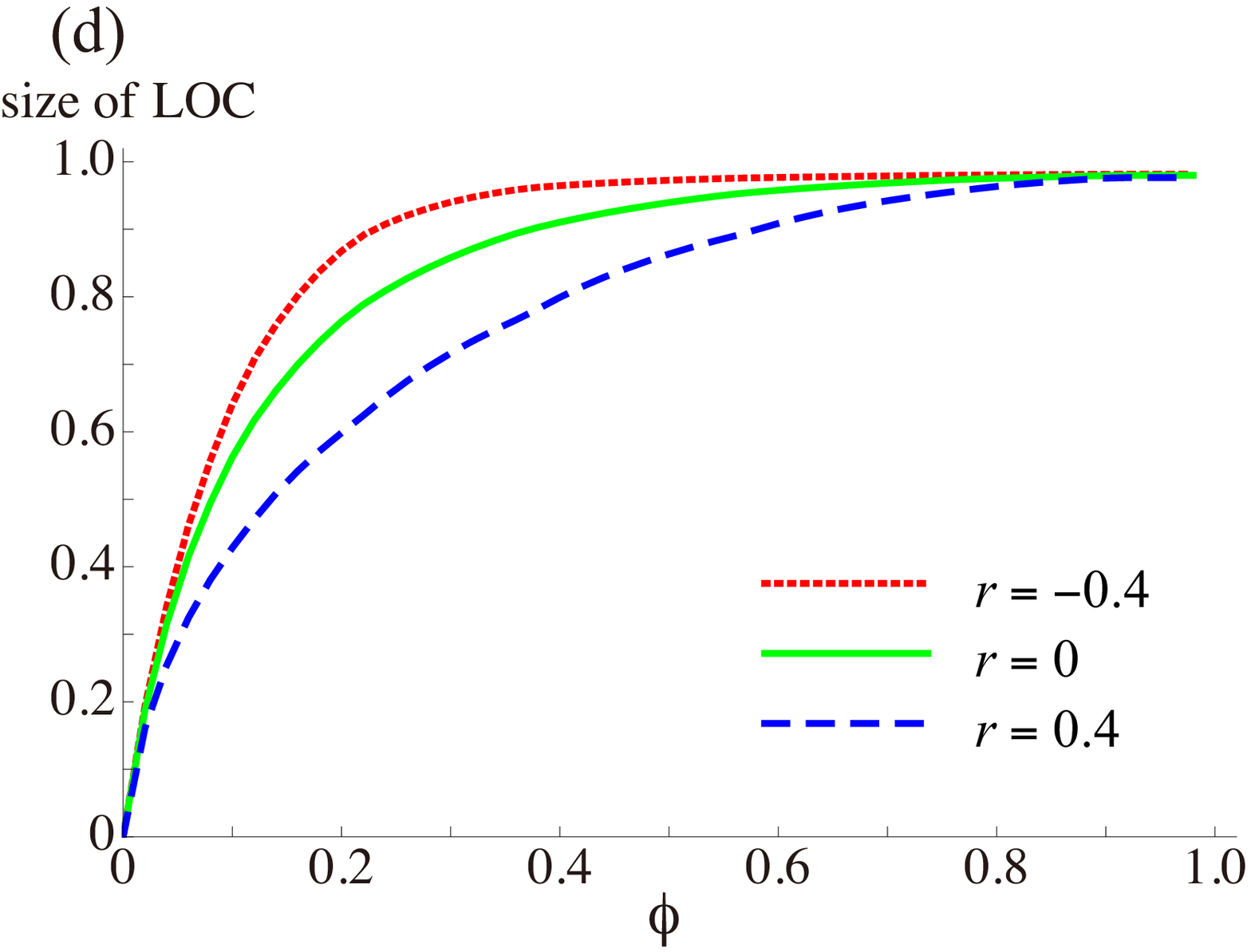}
 \end{center}
 \caption{(Color online) Size of the 
LOC for uncorrelated and correlated networks with $N=10^4$.
(a) Scale-free networks with $p(k)\propto k^{-2.5}$
under random placement.
(b) Scale-free networks with $p(k)\propto k^{-2.5}$
under degree-based placement.
(c) Poisson networks under random placement.
(d) Poisson networks under degree-based placement.
Each data point is an average over $10^3$ realizations of the network on each of which a single placement experiment is carried out.
It should be noted that we set the final $|r|$ values for correlated networks
larger for Poisson networks 
than scale-free networks, because in scale-free networks it is difficult to continue the rewiring as $|r|$ becomes large (e.g., beyond 0.1 or so), whereas the rewiring when $|r|$ is relatively large is much easier for Poisson networks.
}
\label{fig:vary r}
\end{figure}

For the entire range of occupation probability $\phi$, the size of the LOC
normalized by $N$ under the random
placement of sensors is shown in Fig.~\ref{fig:vary r}(a). For each value of $r$ and $\phi$, 
we plotted the size of the LOC averaged over $10^3$ realizations of
networks. On each realization of the network, we carried out a single run of random placement.
Regardless of the degree correlation,
the LOC quickly enlarges as $\phi$ increases. For uncorrelated networks,
this result is consistent with the previous result
\cite{Yang2012}.
Networks with negative degree correlation 
(i.e., $r=-0.1$) and uncorrelated networks (i.e., $r=0$)
realize slightly larger LOCs than
networks with positive degree correlation (i.e., $r=0.1$).

%

Next, we turn to degree-based placement in which hubs
preferentially receive
the sensor.
The size of the LOC under degree-based placement is
shown for scale-free networks with the three $r$ values in Fig.~\ref{fig:vary r}(b). The LOC
for networks with negative (positive) degree correlation
is larger (smaller) than that for uncorrelated networks
for most values of $\phi$. The effect of degree correlation is stronger under degree-based placement (Fig.~\ref{fig:vary r}(b)) than under random placement (Fig.~\ref{fig:vary r}(a)).

The results for Poisson networks under random and degree-based
placement are shown in Figs.~\ref{fig:vary r}(c) and \ref{fig:vary r}(d), respectively. The results are similar to those for scale-free networks (Figs.~\ref{fig:vary r}(a) and \ref{fig:vary r}(b)).

\section{Analysis with generating functions}

\subsection{Uncorrelated networks}\label{sub:uncorrelated GF}

To obtain analytical insights into our numerical results shown in
Sec.~\ref{sub:numerical results}, in this section
we analyze the model by extending the
generating function formalism developed in
Ref.~\cite{Yang2012}.

We start with the case of the general order of
sensor replacement, including random and degree-based placements,
in uncorrelated random and possibly heterogeneous networks.
Denote the probability that a node has degree $k$ by
$p(k)$. The mean degree is given by
$\langle k \rangle = \sum_{k=0}^\infty k p(k)$.
The probability that an endpoint of a randomly selected link
has excess degree, i.e., the number of neighbors minus one, equal to $k$,
is given by
\begin{equation}
q(k) = \frac{(k+1)p(k+1)}{\langle k \rangle}.
\end{equation}
We define the generating functions for the two distributions as
\begin{align}
G_0(x)=& \sum_{k=0}^\infty p(k) x^k,\label{eq:G_0}\\
G_1(x)=& \sum_{k=0}^\infty q(k) x^k.\label{eq:G_1}
\end{align}

We place the sensors on a fraction $\phi$ of nodes in an arbitrary order.
To analyze this situation, we define
\begin{equation}
G_0(x,\rD) = \sum_{k=0}^\infty p(k, \rD) x^k,\label{eq:G_0(D)}
\end{equation}
where $p(k,\rD)$ is the probability that a randomly selected node has degree
  $k$ and is directly observable,
\begin{equation}
G_0(x,\rnD) = \sum_{k=0}^\infty p(k, \rnD) x^k,\label{eq:G_0(nD)}\\
\end{equation}
where $p(k,\rnD)$ is the probability that a randomly selected node has
degree $k$ and is not directly observable (i.e., indirectly
observable or unobservable),
\begin{equation}
G_1(x,\rD) = \sum_{k=0}^\infty q(k, \rD) x^k,\label{eq:G_1(D)}\\
\end{equation}
where $q(k,\rD)$ is the probability that an endpoint of a randomly
selected link has excess degree $k$ and is directly observable, and
\begin{equation}
G_1(x,\rnD) = \sum_{k=0}^\infty q(k, \rnD) x^k,\label{eq:G_1(nD)}
\end{equation}
where $q(k,\rnD)$ is the probability that an endpoint of a randomly
selected link has excess degree $k$ and is not directly observable.
It should be noted that
\begin{equation}
G_0(1,\rD)= 1 - G_0(1,\rnD)=\phi
\end{equation}
and
\begin{equation}
G_1(1,\rD)= 1 - G_1(1,\rnD) = \tphi,
\end{equation}
where $\tphi$ is the probability that an endpoint of a randomly
selected link is directly observable.

For adjacent nodes $i$ and $j$, we recursively calculate two quantities.
First, $s$ is the probability that $i$ does not connect to the LOC through $j$ under the condition that $i$ is indirectly observable
and $j$ is not directly observable (i.e., indirectly observable or unobservable).
Second, $u$ is the probability that $i$ does not connect to the LOC through $j$ under the condition that $i$ is directly observable.

We begin by deriving the recursion equation for $s$.
Assume that $i$ is indirectly observable.
If $j$ is unobservable, $i$ does not connect to the LOC through $j$. This event
occurs with probability
\begin{equation}
\frac{1}{1-\tphi} \sum_{\kj=0}^{\infty} q(\kj,\rnD) (1-\tphi)^{\kj}=\frac{1}{1-\tphi}
G_1(1-\tphi,\rnD).
\label{eq:i=I and j=U}
\end{equation}
Because $q(\kj,\rnD)$ is the  probability that $j$
has excess degree $\kj$ and is not directly observable, we divided
it by $1-\tphi$, i.e., the probability that $j$ is not directly
observable, to derive the conditional probability.
If $j$ is indirectly observable, there are two cases.
First, a neighbor of $j$, denoted by $\ell$,
is directly observable and does not connect to the LOC with probability
$\sum_{\kl=0}^{\infty} q(\kl, \rD) u^\kl = G_1(u, \rD)$. Second, $\ell$ is 
not directly observable
and does not connect to the LOC with probability $\sum_{\kl=0}^{\infty} q(\kl, \rnD) s = G_1(1, \rnD) s = (1-\tphi)s$.
Therefore, if $j$ is indirectly observable, $j$ does not connect to the LOC with probability
\begin{align}
&\frac{1}{1-\tphi} \sum_{\kj=0}^{\infty} q(\kj, \rnD) \sum_{\overline{k}^{\prime}=1}^\kj \binom{\kj}{\overline{k}^{\prime}} G_1(u,
\rD)^{\overline{k}^{\prime}}
\left[(1-\tphi)s\right]^{k-\overline{k}^{\prime}} 
\notag\\
=& \frac{1}{1-\tphi} G_1(\Psi, \rnD)- \frac{1}{1-\tphi} G_1((1-\tphi)s, \rnD),
\label{j=InotLOC}
\end{align}
where
\begin{equation}
\Psi=G_1(u, \rD)+(1-\tphi)s.
\label{eq:Psi uncorrelated}
\end{equation}
It should be noted that, in Eq.~\eqref{j=InotLOC}, the summation over $\overline{k}^{\prime}$ started with
$\overline{k}^{\prime}=1$, not 0, because at least one neighbor of $j$ is
directly observable given that $j$ is indirectly observable.
By adding the right-hand sides of Eqs.~\eqref{eq:i=I and j=U} and \eqref{j=InotLOC}, we obtain
\begin{equation}
s= \frac{1}{1-\tphi} G_1(1-\tphi,\rnD) + \frac{1}{1-\tphi} G_1(\Psi,\rnD) -\frac{1}{1-\tphi} G_1((1-\tphi)s,\rnD). 
\label{eq:s uncorrelated}
\end{equation}

To derive the recursion equation for $u$, we
assume that $i$ is directly observable.
The probability that $j$ is directly observable and $i$ does not
connect to the LOC through $j$ is equal to $\sum_{\kj=0}^{\infty} q(\kj, \rD) u^\kj =
G_1(u, \rD)$. Otherwise, $j$ is indirectly observable because it is
adjacent to $i$, which is directly observable.
In the latter case, $i$ does not connect to the
LOC through $j$ with probability $G_1(\Psi, \rnD)$.
Therefore, we obtain
\begin{equation}
u=G_1(u, \rD)+G_1(\Psi, \rnD).
\label{eq:u uncorrelated}
\end{equation}

Node $i$ is directly observable and belongs to the LOC with
probability
\begin{equation}
\sum_{k=1}^{\infty} p(k, \rD) (1-u^k)=\phi -G_0(u, \rD).
\label{eq:i=D and belongs to LOC}
\end{equation}
Node $i$ is indirectly observable and belongs to the LOC with probability
\begin{align}
&\sum_{k=1}^{\infty} p(k, \rnD) \sum_{\overline{k}=1}^k \binom{k}{\overline{k}}\tphi^{\overline{k}} (1-\tphi)^{k-\overline{k}} \left[1-\left(\frac{G_1(u, \rD)}{\tphi} \right)^{\overline{k}} s^{k-\overline{k}} \right] \notag\\
=&1-\phi - G_0(\Psi, \rnD) - G_0(1-\tphi, \rnD) + G_0((1-\tphi)s, \rnD).
\label{eq:i=I and belongs to LOC}
\end{align}
It should be noted that
$G_1(u, \rD)/\tphi$ is the probability that node $j$ adjacent to node
$i$ does not connect to the LOC via any neighbor of $j$, given that $j$ is directly observable.
Therefore, the probability that a randomly selected node belongs to
the LOC, which is identified by the size of the LOC, denoted by $S$,
is given by the summation of the right-hand sides of Eqs.~\eqref{eq:i=D and belongs to LOC} and \eqref{eq:i=I and belongs to LOC}, i.e.,
\begin{equation}
S=1 - G_0(u, \rD) - G_0(\Psi, \rnD) - G_0(1-\tphi, \rnD) + G_0((1-\tphi)s, \rnD). 
\label{eq:S uncorrelated}
\end{equation}

In the case of random replacement,
we set
$p(k, \rD)=\phi p(k)$, $p(k, \rnD)=(1-\phi) p(k)$,
$q(k, \rD)=\phi q(k)$, $q(k, \rnD)=(1-\phi) q(k)$,
$G_0(x, \rD)=\phi G_0(x)$, $G_0(x,\rnD)=(1-\phi)G_0(x)$,
$G_1(x, \rD)=\phi G_1(x)$, and $G_1(x,\rnD)=(1-\phi)G_1(x)$
to obtain $\tphi=\phi$.
Then, Eqs.~\eqref{eq:Psi uncorrelated}, \eqref{eq:s uncorrelated}, \eqref{eq:u uncorrelated}, and \eqref{eq:S uncorrelated} are reduced to
\begin{align}
S=& 1 - \phi G_0(u) - (1-\phi)\left[G_0(\Psi)+G_0(1-\phi)-G_0((1-\phi)s)\right],\\
s =& G_1(1-\phi) + G_1(\Psi) -G_1((1-\phi)s),\\
u =& \phi G_1(u)+(1-\phi )G_1(\Psi),
\end{align}
where
\begin{equation}
\Psi=\phi G_1(u)+(1-\phi)s.
\end{equation}
These results agree with those in Ref.~\cite{Yang2012}.

In the case of degree-based placement, we set
\begin{align}
p(k, \rD) =& 
\begin{cases}
	p(k) & (k>k_{\rm cut}),\\
	\alpha p(k) & (k=k_{\rm cut}),\\
	0 & (k<k_{\rm cut}),
\end{cases}
\label{eq:p(k,D) degree-based}\\
p(k, \rnD) =& 
\begin{cases}
	0 & (k>k_{\rm cut}),\\
	(1-\alpha) p(k) & (k=k_{\rm cut}),\\
	p(k) & (k<k_{\rm cut}),
\end{cases}
\label{eq:p(k,nD) degree-based}\\
q(k, \rD) =& 
\begin{cases}
	q(k) & (k>k_{\rm cut}),\\
	\alpha q(k) & (k=k_{\rm cut}),\\
	0 & (k<k_{\rm cut}),
\end{cases}
\label{eq:q(k,D) degree-based}\\
q(k, \rnD) =& 
\begin{cases}
	0 & (k>k_{\rm cut}),\\
	(1-\alpha) q(k) & (k=k_{\rm cut}),\\
	q(k) & (k<k_{\rm cut}),
\end{cases}
\label{eq:q(k,nD) degree-based}
\end{align}
where $k_{\rm cut}$ is the minimum degree of the node at which the
sensor is placed.
The fraction of nodes with degree
$k=k_{\rm cut}$ that receive sensors, denoted by $\alpha$,
 is determined by
\begin{equation}
\sum_{k=k_{\rm cut}+1}^\infty p(k)+\alpha p(k_{\rm cut})=\phi.
\end{equation}
We obtain $S$ by substituting Eqs.~\eqref{eq:p(k,D)
  degree-based}--\eqref{eq:q(k,nD) degree-based} in
Eqs.~\eqref{eq:G_0(D)}--\eqref{eq:G_1(nD)} and calculating
Eqs.~\eqref{eq:s uncorrelated}, \eqref{eq:u uncorrelated}, and
\eqref{eq:S uncorrelated} with 
\begin{equation}
\tphi=\sum_{k=k_{\rm cut}+1}^\infty q(k)+\alpha q(k_{\rm cut}).
\end{equation}

\begin{figure}[!h]
 \begin{center}
  \includegraphics[width=75mm]{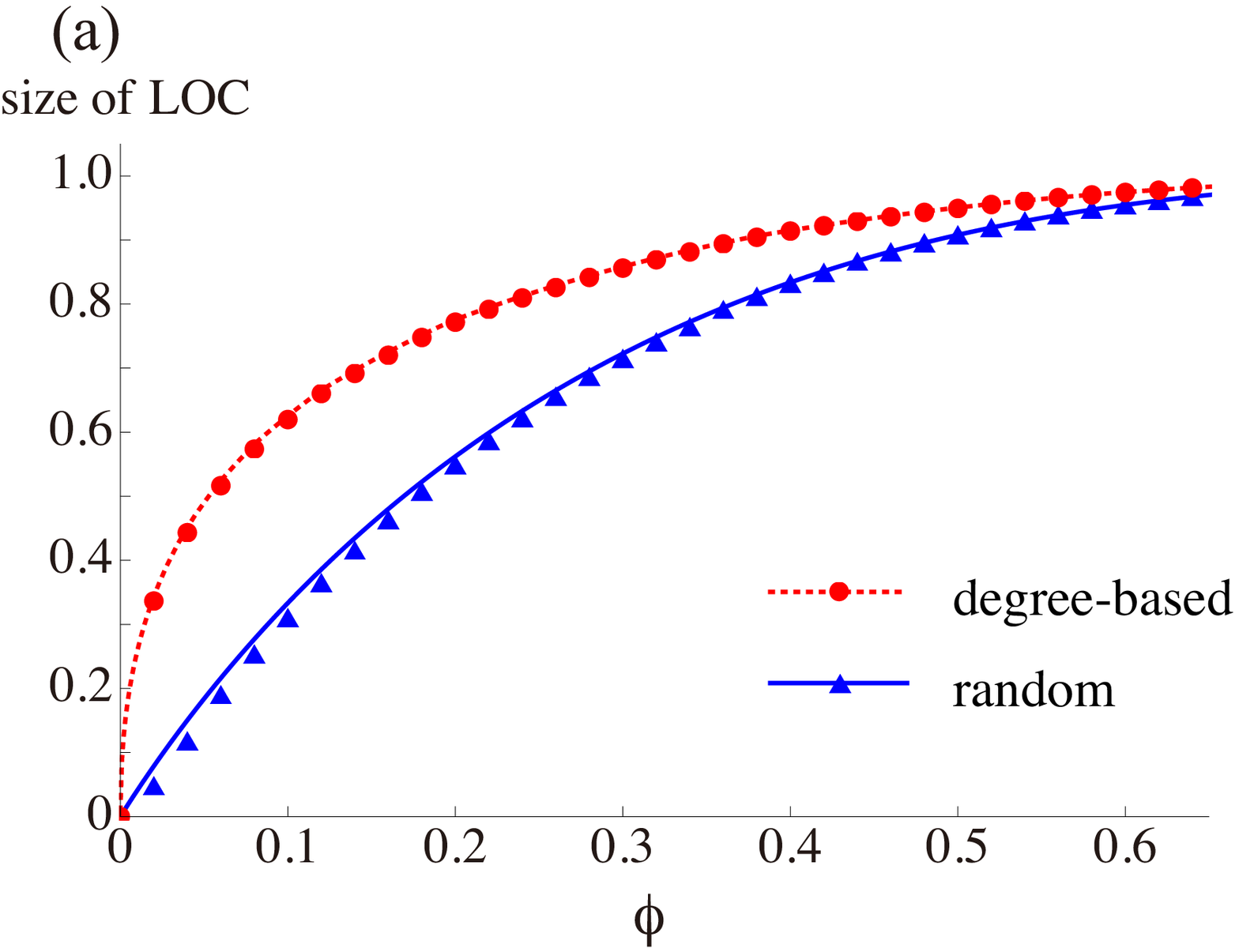}
  \includegraphics[width=75mm]{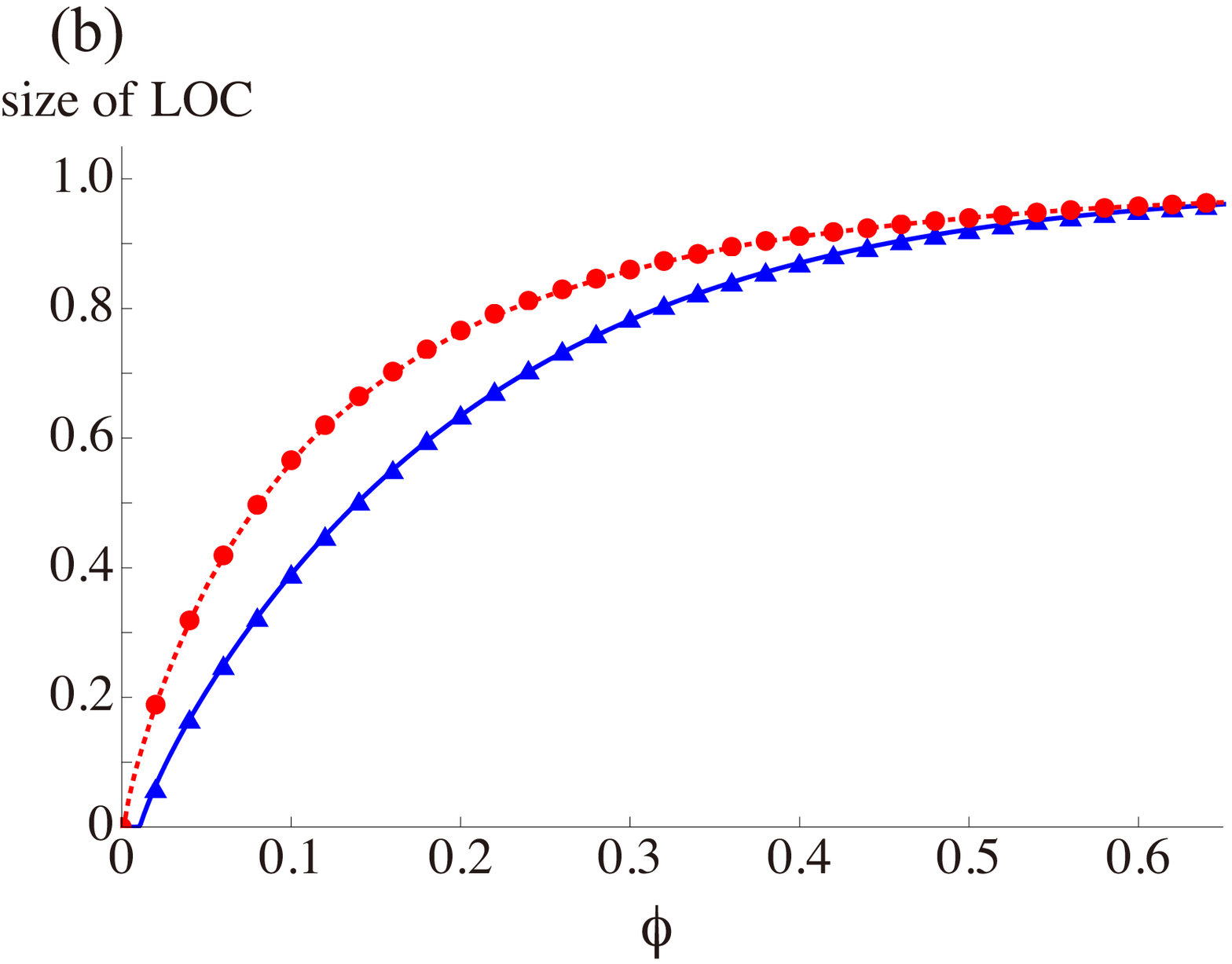}
 \end{center}
 \caption{(Color online) Size of the LOC for random and degree-based placements on
uncorrelated networks with $N=10^4$ and $\left<k\right>\approx 4$.
(a) Scale-free networks generated by the configuration model with $p(k)\propto
k^{-3}$ ($k\ge 2$). (b) ER random graph. 
The triangles and circles represent the numerical results for
random and degree-based placements, respectively.
Each data point is an average value on the basis of 10 realizations of the network. The solid and dotted lines represent the theoretical results for
random and degree-based placements, respectively.}
\label{fig:uncorrelated with GF}
\end{figure}

The LOC for the uncorrelated 
scale-free networks with $p(k)\propto k^{-3}$, where the minimum degree is set to 2, and the ER random graph are shown in
Figs.~\ref{fig:uncorrelated with GF}(a) and
\ref{fig:uncorrelated with GF}(b), respectively.
For both random and
degree-based placements, the analytical results on the basis of the
generation functions (lines)
agree very well with
those obtained from numerical simulations (symbols).
The results for random placement replicate those in Ref.~\cite{Yang2012}.
Degree-based placement only slightly shifts
the critical $\phi$ value at which the observability transition
occurs. This is mainly because the critical $\phi$ value
is small even for random
placement \cite{Yang2012}. This is in particular the case for the scale-free networks
(Fig.~\ref{fig:uncorrelated with GF}(a)).
However, for both networks, 
the degree-based placement increases the size of the LOC
across a wide range of $\phi$ as compared to the case of random placement.

\subsection{Correlated networks}

In this section, we generalize the theory
developed in Sec.~\ref{sub:uncorrelated GF} to the case
of networks with degree correlation.
To this end, denote by 
$p(k,\kj)$ the probability that the two nodes
adjacent to each other through a randomly selected link have degrees
$k$ and $\kj$. The normalization is given by
$\sum_{k=1}^{\infty}\sum_{\kj=1}^{\infty}p(k,\kj)=1$. Denote by
$p(\kj|k)$ the probability that a
neighbor of a node with degree $k$ has degree $\kj$.
It should be noted that $\sum_{\kj=1}^{\infty} p(\kj|k)=1$ and
that $\left[kp(k)/\langle k \rangle\right] p(\kj|k)=
p(k,\kj)=p(\kj,k)=\left[\kj p(\kj)/\langle k \rangle\right] p(k|\kj)$~\cite{Boguna2002,Serrano2007}. 

As in Sec.~\ref{sub:uncorrelated GF}, we assume that
a fraction $\phi$ of nodes
selected according to a given order is
directly observable. In addition to
$p(k,\rD)$ and $p(k,\rnD)$,
we use $p(\kj,\rD | k)$ and $p(\kj,\rnD | k)$, which
are defined as the probabilities that
a neighbor of a node with degree $k$ has degree $\kj$ and is directly
observable and not directly observable, respectively.
Assume that a node $i$ with degree $k$ and a node $j$ with
degree $\kj$ are adjacent to each other. We are going to calculate
$s_k$, the probability that $i$ is
not connected to the LOC through $j$ under the condition that
$i$ is indirectly observable and $j$ is not directly observable (i.e.,
indirectly observable or unobservable), and 
$u_k$, the probability that $i$ is
not connected to the LOC through $j$ under the condition that
$i$ is directly observable.

Assume that $i$ is indirectly observable.
We denote by $\kl$ the degree of 
node $\ell$, which is a neighbor of
$j$ other than $i$. These notations (i.e., $\kl$ and $\ell$) are consistent with those we used
in Sec.~\ref{sub:uncorrelated GF}.
The contribution to $s_k$ of the case in which $j$ is unobservable
is equal to
\begin{equation}
\frac{1}{\sum_{\kj=1}^{\infty} p(\kj, \rnD | k)} \sum_{\kj=1}^{\infty} p(\kj, \rnD | k)
\left[\sum_{\kl=1}^{\infty} p(\kl, \rnD | \kj) \right]^{\kj-1}.
\label{eq:i=I j=U correlated}
\end{equation}
When $j$ is indirectly observable, there are two cases.
First, $\ell$ is directly observable 
and does not belong to the LOC with probability
$\sum_{\kl=1}^{\infty} p(\kl, \rD | \kj)(u_\kl)^{\kl-1}$.
Second, $\ell$ is not directly observable and does not belong
to the LOC with probability $\sum_{\kl=1}^{\infty} p(\kl, \rnD | \kj)s_{k^{\prime}}$.
Therefore, given that $j$ is indirectly observable, it does not belong to
the LOC with probability
\begin{align}
& \frac{1}{\sum_{\kj=1}^{\infty} p(\kj, \rnD | k)} \sum_{\kj=1}^{\infty} p(\kj, \rnD | k) \sum_{\overline{k}^{\prime}=1}^{\kj-1} \binom{\kj-1}{\overline{k}^{\prime}} \left[\sum_{\kl=1}^{\infty} p(\kl, \rD | \kj) (u_\kl)^{\kl-1}\right]^{\overline{k}^{\prime}} \left[\sum_{\kl=1}^{\infty} p(\kl, \rnD | \kj) s_\kj\right]^{\kj-1-\overline{k}^{\prime}}\notag\\
=& \frac{1}{\sum_{\kj=1}^{\infty} p(\kj, \rnD | k)} \left\{ \sum_{\kj=1}^{\infty} p(\kj, \rnD | k) (\Psi_\kj)^{\kj-1} - \sum_{\kj=1}^{\infty} p(\kj, \rnD | k) \left[\sum_{\kl=1}^{\infty} p(\kl, \rnD | \kj)s_\kj \right]^{\kj-1}\right\},
\label{eq:i=I j=I correlated}
\end{align}
where
\begin{equation}
\Psi_\kj=\sum_{\kl=1}^{\infty} p(\kl, \rD | \kj) (u_\kl)^{\kl-1} + \sum_{\kl=1}^{\infty} p(\kl, \rnD | \kj)s_\kj.
\label{eq:Psi_k correlated}
\end{equation}
Using Eqs.~\eqref{eq:i=I j=U correlated}
and \eqref{eq:i=I j=I correlated}, we obtain
\begin{align}
s_k =& \frac{1}{\sum_{\kj=1}^{\infty} p(\kj, \rnD | k)} 
\left\{ \sum_{\kj=1}^{\infty} p(\kj, \rnD | k) (\Psi_\kj)^{\kj-1}+\sum_{\kj=1}^{\infty} p(\kj, \rnD | k) \left[\sum_{\kl=1}^{\infty} p(\kl, \rnD | \kj) \right]^{\kj-1} \left[1-(s_\kj)^{\kj-1}\right] \right\}.
\label{eq:s_ki}
\end{align}

Next, we obtain
\begin{equation}
u_k=\sum_{\kj=1}^{\infty} p(\kj, \rD | k) (u_\kj)^{\kj-1}+\sum_{\kj=1}^{\infty} p(\kj, \rnD | k) (\Psi_\kj)^{\kj-1}.
\label{eq:u_ki}
\end{equation}
The first term on the right-hand side of Eq.~\eqref{eq:u_ki}
is the probability that
$j$ is directly observable and does not belong to the LOC.
The second term is the probability that
$j$ is indirectly observable and does not belong to the LOC, i.e.,
$\sum_{\kj=1}^{\infty} p(\kj, \rnD | k) \left[\sum_{\kl=1}^{\infty} p(\kl, \rD | \kj)
(u_\kl)^{\kl-1} + \sum_{\kl=1}^{\infty} p(\kl, \rnD | \kj) s_\kj\right]^{\kj-1}$.

Node $i$ is directly observable and belongs to the LOC with probability
\begin{equation}
\sum_{k=1}^{\infty} p(k, \rD) \left[1-(u_k)^k\right]=\phi - \sum_{k=0}^{\infty} p(k, \rD) (u_k)^k.
\label{eq:i=D LOC correlated}
\end{equation}
Node $i$ is indirectly observable (and hence at least one neighbor of $i$
is directly observable) and belongs to the LOC with probability
\begin{align}
&\sum_{k=1}^{\infty} p(k, \rnD) \sum_{\overline{k}=1}^k \binom{k}{\overline{k}} \left[\sum_{\kj=1}^{\infty} p(\kj, \rD|k) \right]^{\overline{k}} \left[\sum_{\kj=1}^{\infty} p(\kj, \rnD|k) \right]^{k-\overline{k}} \left\{1-\left[\frac{\sum_{\kj=1}^{\infty} p(\kj, \rD|k) (u_\kj)^{\kj-1}}{\sum_{\kj=1}^{\infty} p(\kj, \rD|k)} \right]^{\overline{k}} (s_k)^{k-\overline{k}} \right\}\notag\\
=& 1-\phi -\sum_{k=0}^{\infty} p(k, \rnD) \left[\sum_{\kj=1}^{\infty} p(\kj, \rD|k) (u_\kj)^{\kj-1}+\sum_{\kj=1}^{\infty} p(\kj, \rnD|k) s_k \right]^k 
- \sum_{k=0}^{\infty} p(k, \rnD) \left[ \sum_{\kj=1}^{\infty} p(\kj, \rnD|k) \right]^k \left[1-(s_k)^k\right]\notag\\
=&1-\phi -\sum_{k=0}^{\infty} p(k, \rnD) (\Psi_k )^k - \sum_{k=0}^{\infty} p(k, \rnD) \left[ \sum_{\kj=1}^{\infty} p(\kj, \rnD|k) \right]^k \left[1-(s_k)^k\right].
\label{eq:i=I LOC correlated}
\end{align}
By summing the right-hand sides of Eqs.~\eqref{eq:i=D LOC correlated}
and \eqref{eq:i=I LOC correlated}, we obtain
\begin{equation}
S=1-\sum_{k=0}^{\infty} p(k, \rD) (u_k)^k 
-\sum_{k=0}^{\infty} p (k, \rnD) (\Psi_k)^k 
-\sum_{k=0}^{\infty} p (k, \rnD) \left[\sum_{\kj=1}^{\infty} p(\kj, \rnD | k) \right]^k
\left[1-(s_k)^k \right].
\label{LOC-target}
\end{equation}

In the case of random replacement,
we use $p(k, \rD)= \phi p(k)$, $p(k, \rnD)=(1-\phi) p(k)$,
$p(\kj, \rD|k)= \phi p(\kj|k)$, $p(\kj, \rnD |k)=(1-\phi) p(\kj|k)$,
$\sum_{\kj=1}^{\infty} p(\kj, \rD| k)=\phi \sum_{\kj=1}^{\infty} p(\kj|k)=\phi$, and $\sum_{\kj=1}^{\infty}
p(\kj, \rnD| k)=(1-\phi) \sum_{\kj=1}^{\infty} p(\kj|k)=1-\phi$
to simplify
Eqs. \eqref{eq:Psi_k correlated}, \eqref{eq:s_ki}, \eqref{eq:u_ki}, and \eqref{LOC-target} as
\begin{equation}
S=1- \phi \sum_{k=0}^{\infty} p(k) (u_k)^k -(1-\phi) \left\{\sum_{k=0}^{\infty} p(k) (\Psi_k)^k + \sum_{k=0}^{\infty} p(k) (1-\phi)^k \left[1-(s_k)^k\right] \right\},
\label{eq:S correlated net, random replacement}
\end{equation}
where
\begin{align}
s_k =& \sum_{\kj=1}^{\infty} p (\kj | k) (\Psi_\kj)^{\kj-1}+ \sum_{\kj=1}^{\infty}
p(\kj | k) (1-\phi)^{\kj-1} \left[1-(s_\kj)^{\kj-1}\right],
\label{eq:s_ki correlated net, random replacement}
\\
u_k =& \phi \sum_{\kj=1}^{\infty} p(\kj | k) (u_\kj)^{\kj-1} + (1-\phi) \sum_{\kj=1}^{\infty} p(\kj | k) (\Psi_\kj)^{\kj-1},
\label{eq:u_ki correlated net, random replacement}
\end{align}
and
\begin{equation}
\Psi_\kj = \phi \sum_{\kl=1}^{\infty} p(\kl | \kj) (u_\kl)^{\kl-1} + (1-\phi) s_\kj.
\label{eq:Psi_kj correlated net, random replacement}
\end{equation}

The results for correlated networks obtained so far involve
many auxiliary variables, i.e., $s_k$ and $u_k$. It seems difficult
to simplify the results even for the case of random placement
(Eqs.~\eqref{eq:S correlated net, random
  replacement}--\eqref{eq:Psi_kj correlated net, random
  replacement}). 
Therefore, in the rest of this section,
we confine ourselves to bimodal
networks in which there are only two degree values.
Assume that
$Na$ and $N(1-a)$ nodes have degrees $k_1$ and $k_2 (<k_1)$, respectively,
such that 
\begin{equation}
p(k) = a \delta_{k,k_1}+(1-a) \delta_{k,k_2}
\end{equation}
and
\begin{equation}
\left<k\right> = \sum_{k=k_1, k_2} kp(k)=a k_1+ (1-a) k_2,
\end{equation}
where $\delta$ is the Kronecker delta.
Given a conditional degree distribution
$p(k_1|k_2) (=1-p(k_2|k_2))$, the other conditional degree distribution is
automatically determined as
$p(k_2|k_1)=k_2 (1-a)p(k_1|k_2)/a k_1
(=1-p(k_1|k_1))$. The degree correlation is equal to
\begin{equation}
r = 1-\frac{\left[a k_1+(1-a) k_2\right]p(k_1|k_2)}{a k_1}.
\label{eq:r bimodal}
\end{equation}

In the case of random placement, we use
Eqs.~\eqref{eq:S correlated net, random
  replacement}--\eqref{eq:Psi_kj correlated net, random
  replacement}, which involve four recursively calculated variables
$s_{k_1}$, $s_{k_2}$, $u_{k_1}$, and $u_{k_2}$.
In the case of degree-based placement, for $0\le \phi\le a$, we
use $p(k_1, \rD) = \phi$,
$p(k_1, \rnD) = a-\phi$,
$p(k_2, \rD) = 0$,
$p(k_2, \rnD) = 1-a$,
$p(k_1, \rD |k_1) =  p(k_1|k_1)\phi/a$,
$p(k_1, \rnD |k_1) = p(k_1|k_1)\left[1- (\phi/a) \right]$,
$p(k_2, \rD |k_1) = 0$,
$p(k_2, \rnD |k_1) = p(k_2|k_1)$,
$p(k_1, \rD |k_2) = p(k_1|k_2)\phi/a$,
$p(k_1, \rnD |k_2) = p(k_1|k_2)\left[1- \phi/a \right]$,
$p(k_2, \rD |k_2) = 0$, and
$p(k_2, \rnD |k_2) = p(k_2|k_2)$.
We substitute these equations in
Eqs.~\eqref{eq:Psi_k correlated}, \eqref{eq:s_ki}, \eqref{eq:u_ki}, and \eqref{LOC-target} to obtain $S$ for
$0\le \phi\le a$.
For $a<\phi\le 1$, we use
$p(k_1, \rD) = a$,
$p(k_1, \rnD) = 0$,
$p(k_2, \rD) = \phi-a$,
$p(k_2, \rnD) = 1-\phi$,
$p(k_1, \rD |k_1) = p(k_1|k_1)$,
$p(k_1, \rnD |k_1) = 0$,
$p(k_2, \rD |k_1) = p(k_2|k_1) (\phi-a)/(1-a)$,
$p(k_2, \rnD |k_1) = p(k_2|k_1) (1-\phi)/(1-a)$,
$p(k_1, \rD |k_2) = p(k_1|k_2)$,
$p(k_1, \rnD |k_2) = 0$,
$p(k_2, \rD |k_2) = p(k_2|k_2) (\phi-a)/(1-a)$, and
$p(k_2, \rnD |k_2) = p(k_2|k_2) (1-\phi)/(1-a)$.

\begin{figure}[!h]
 \begin{center}
\includegraphics[width=75mm]{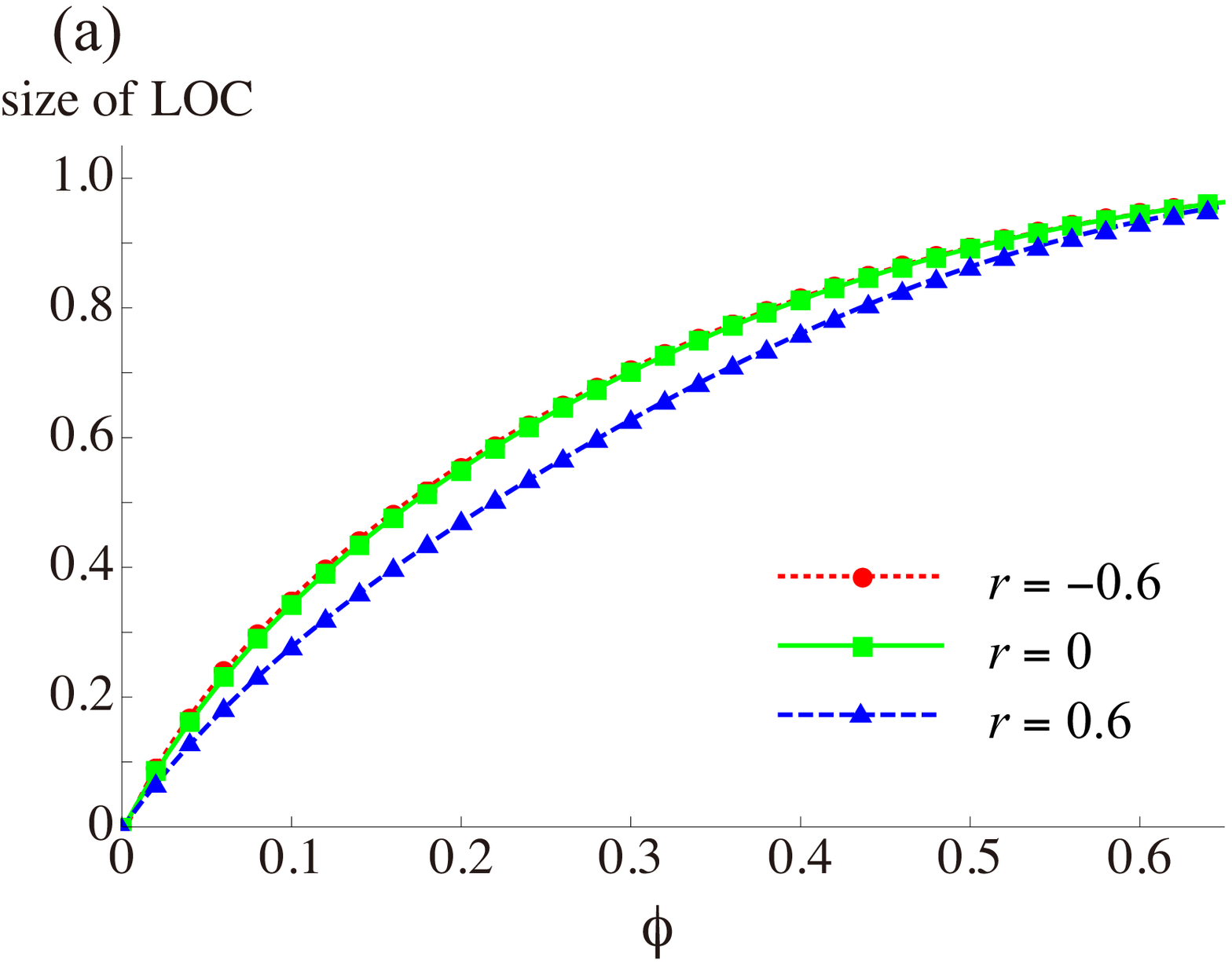}
\includegraphics[width=75mm]{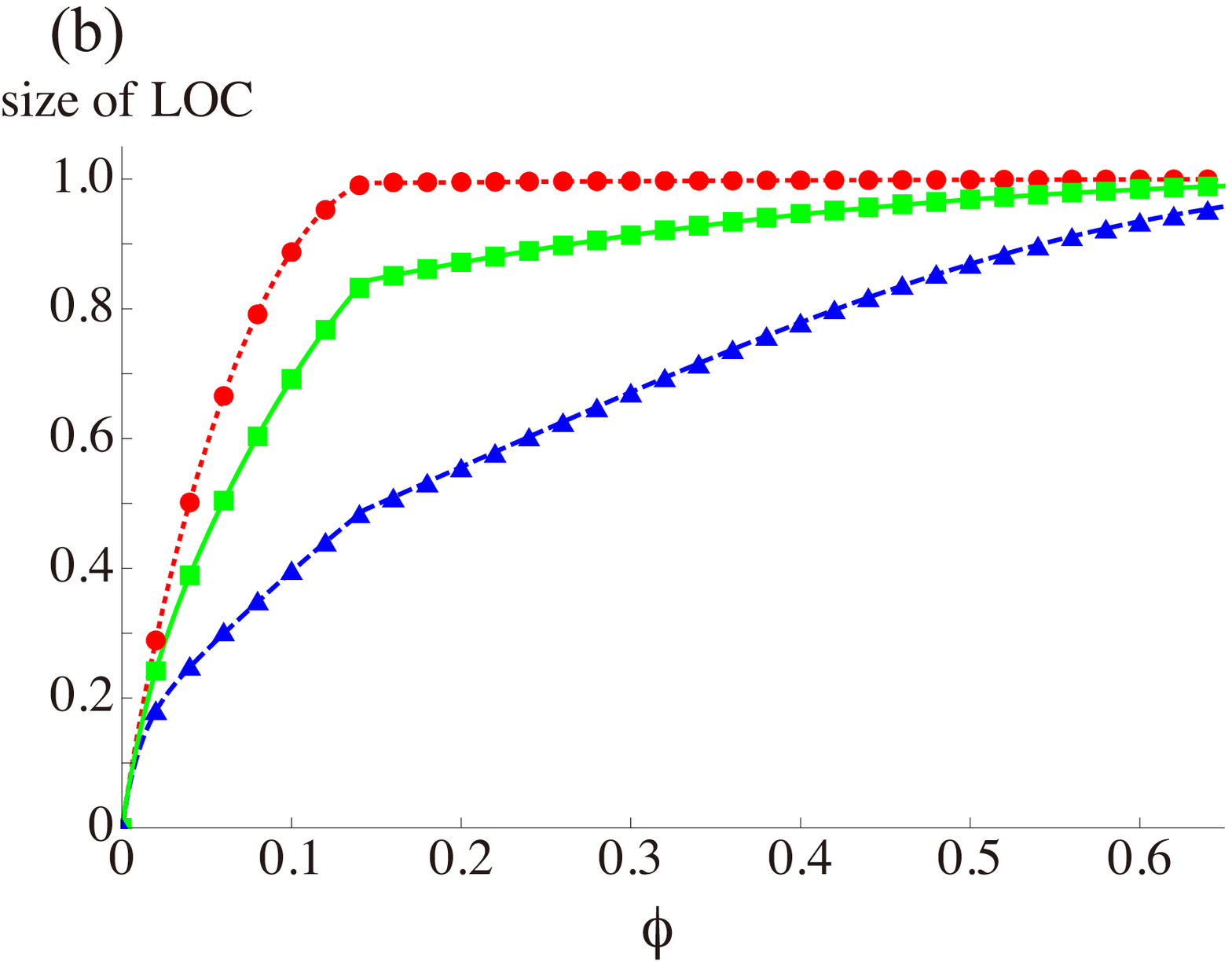}
 \end{center}
 \caption{(Color online) LOC size for bimodal networks with $k_1=16$, $k_2=2$, and $a=1/7$.
(a) Random placement. (b) Degree-based
placement. The dotted, solid, and dashed lines represent the theoretical
results obtained with the generation functions 
for $r=-0.6$, $r=0$, and $r=0.6$, respectively.
The circles, squares, and triangles represent the numerical results
for $r=-0.6$, $r=0$, and $r=0.6$, respectively.
}
 \label{fig:bimodal}
\end{figure}

The size of the LOC for uncorrelated and correlated bimodal networks
is shown in
Fig.~\ref{fig:bimodal}. 
Because we set  $k_1=16$, $k_2=2$, and $a=1/7$, 
we obtain $\left<k\right>=4$. By combining Eq.~\eqref{eq:r bimodal}
and $0<p(k_1|k_2)<1$, we obtain
$-3/4 < r< 1$. We set $r=-0.6$, 0, and 0.6
in Fig.~\ref{fig:bimodal}.
In the figure, the numerical results are shown by circles, squares, and triangles. They are calculated for random bimodal networks possessing
$N=10^4$ nodes, and the size of the LOC at each $\phi$ value
is an average over 10 realizations of networks.
For both random placement (Fig.~\ref{fig:bimodal}(a)) and degree-based placement (Fig.~\ref{fig:bimodal}(b)),
the numerical results agree very well
with the analytical results derived from the generating functions
shown by the lines.
The effect of the degree correlation on the size of the LOC is small
under random replacement. In contrast, the degree correlation
considerably changes the size of the LOC
under degree-based placement.
However, the critical $\phi$ value does not depend much on the degree
correlation even in the case of degree-based placement.
These results are qualitatively the same as those for
scale-free and Poisson networks shown in Fig.~\ref{fig:vary r}.
 It should be noted that the increase in the size of the LOC is not smooth at $\phi=a=1/7$ under degree-based placement (most evident for $r=0$; solid line in Fig.~\ref{fig:bimodal}(b)) because the analytical expression for the LOC size changes at $\phi=a$.

\section{Networks optimized for the size of the LOC}

To support our claim that negative degree correlation enlarges
the observable component,
we numerically explore optimal network structure for observability transitions
under degree-based placement.
The results shown in the previous sections indicate
that degree-based
placement considerably enhances the size of the LOC for a wide range of $\phi$, whereas 
it does not change the critical $\phi$ value much. Therefore, similar to the analysis of the percolation transition~\cite{Herrmann2011,Schneider2011,Tanizawa2012},
we define the objective function to be optimized
as the size of the LOC summed over $\phi$ values, i.e.,
\begin{equation}
R_{\rm O} = \frac{1}{N+1} \sum_{i=0}^N O(i),
\label{eq:R_0}
\end{equation}
where $O(i)$ ($0\le O(i)\le 1$) is the relative size of the LOC when the
sensors are placed on the first $i (=N\phi)$ nodes in the descending order of degree. It should be noted that $O(0)=0$ and $O(1)=1$.

Herrmann and colleagues used an optimization method to generate the most tolerant network against the targeted (i.e., degree-based) attack~\cite{Herrmann2011,Schneider2011} (also see \cite{Tanizawa2012}). They optimized a given network
by repetitively rewiring
two randomly selected links without changing the degree of each node,
if the rewiring increased a tolerance measure.
We carry out a similar optimization experiment as follows.
First, we prepare uncorrelated
networks according to the configuration model or the ER random
graph. Second, we select two links at random and tentatively rewire them
in the same manner as the method for generating
correlated networks (Sec.~\ref{sub:generation corr
  nets}). If the tentative rewiring increases $R_{\rm O}$, we adopt
the rewiring if it does not yield multiple links
or self-loops. Otherwise, we
discard the proposed rewiring. We repeat the tentative rewiring, including
unsuccessful attempts, $10^7$ times.
We set $N=10^4$.

The $R_{\rm O}$ values before and after
the optimization of the scale-free network with $p(k)\propto
k^{-2.5}$ and $\left<k\right>\approx 4$ are equal to
0.9178 $\pm$ 0.0015 and 0.9943 $\pm$ 0.0003 (mean $\pm$ standard deviation), respectively, where
the statistical values are calculated on the basis of 10 realizations of networks.
The optimization procedure considerably increases
the size of the LOC. The results are qualitatively the same for
Poisson networks. The $R_{\rm O}$ values before and after the optimization are equal to 0.8493 $\pm$ 0.0014 and 0.9247 $\pm$ 0.0011, respectively.

The optimization in terms of the LOC does not enlarge
the conventional percolation cluster. To show this, we measure the order parameter
$R_{\rm D} \equiv \sum_{i=0}^N D(i)/(N+1)$,
where $D(i)$ is the normalized size of the largest connected component (LCC) when the $i$ nodes are
occupied in the descending order of degree of the original
network. By definition, $R_{\rm D}$ is large if the LCC composed of the directly observable nodes is large for a wide range of $\phi$.
$R_{\rm D}$ is analogous to the one previously used
for analyzing targeted attacks~\cite{Herrmann2011,Schneider2011,Tanizawa2012} and summarizes the size of the LCC across various values of $\phi$.
If $R_{\rm D}$ is large, placing the sensor on a relatively small number of nodes can make the LCC large. 
Throughout the optimization procedure in terms of the LOC described above,
$R_{\rm D}$ changes from 0.4937 $\pm$ 0.0005 to 0.4990 $\pm$ 0.0001 for scale-free networks.
The size of the LCC is close to 0.5, which is 
the maximum possible value realized when $D(i) = i/N$,
even before the optimization. Partly for this reason, the increase in $R_{\rm D}$ is small albeit significant. For Poisson networks, $R_{\rm D}$ significantly decreases from
0.4909 $\pm$ 0.0007 to 0.4722 $\pm$ 0.0003 through the optimization in terms of $R_{\rm O}$. 

For scale-free networks with different $\gamma$ values and the ER random graph,
the degree correlation after the optimization
is shown in Fig.~\ref{fig:optimize vary
  gamma} (circles). The $r$ values corresponding to $\gamma=\infty$ indicate the
results for the ER random graph. The optimized networks
have negative $r$ values
irrespective of the degree distribution. This result
is consistent with those obtained in Sec.~\ref{sub:numerical results}.
The negative $r$ value is not caused by
the fact that 
scale-free networks with small $\gamma$ have inherently negative $r$ values
~\cite{Maslov2004,Park2003,Boguna2004,Catanzaro2005,Alderson2007,Weber2007,Serrano2007}; we confirmed that
$r$ did decrease as a result of optimization for any $\gamma$.

\begin{figure}[!h]
 \begin{center}
  \includegraphics[width=75mm]{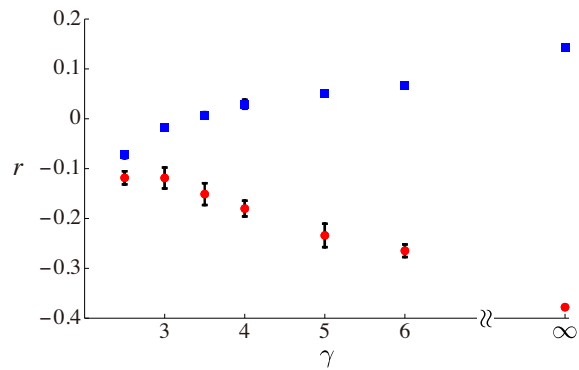}
 \end{center}
 \caption{(Color online) Degree correlation after the optimization. We set
   $N=10^4$. The circles and squares correspond to the results for
   the optimization in terms of $R_{\rm O}$ and $R_{\rm D}$, respectively.
We denote by $\gamma$ the scale-free exponent of the degree distribution
(i.e., $p(k)\propto k^{-\gamma}$), and $\gamma=\infty$ corresponds to Poisson networks.
Error bars are the standard deviations on the basis of
10 realizations. They are smaller than the size of the circle when $\gamma=\infty$ and the squares for all the $\gamma$ values.}
\label{fig:optimize vary gamma}
\end{figure}

To contrast the negative degree correlation emerging as a result of the optimization in terms of $R_{\rm O}$, we also carried out another set of numerical optimization experiments. We optimized the scale-free and Poisson networks
in terms of $R_{\rm D}$.
The $r$ values after the optimization,
shown by the squares in Fig.~\ref{fig:optimize vary
  gamma}, have different signs depending on
the degree distribution.
In general, highly heterogeneous networks
tend to produce negative $r$ even if they are constructed from the configuration model~\cite{Maslov2004,Park2003,Boguna2004,Catanzaro2005,Alderson2007,Weber2007,Serrano2007}. The theory suggests that this phenomenon occurs only for $\gamma < 3$ \cite{Boguna2004}.
These previous results are consistent with the results shown in
 Fig.~\ref{fig:optimize vary gamma}. Therefore, we conclude that
the LCC tends to be large with 
positive degree correlation, whereas the LOC tends to be large with negative degree correlation.

\begin{figure}[!h]
 \begin{center}
  \includegraphics[width=75mm]{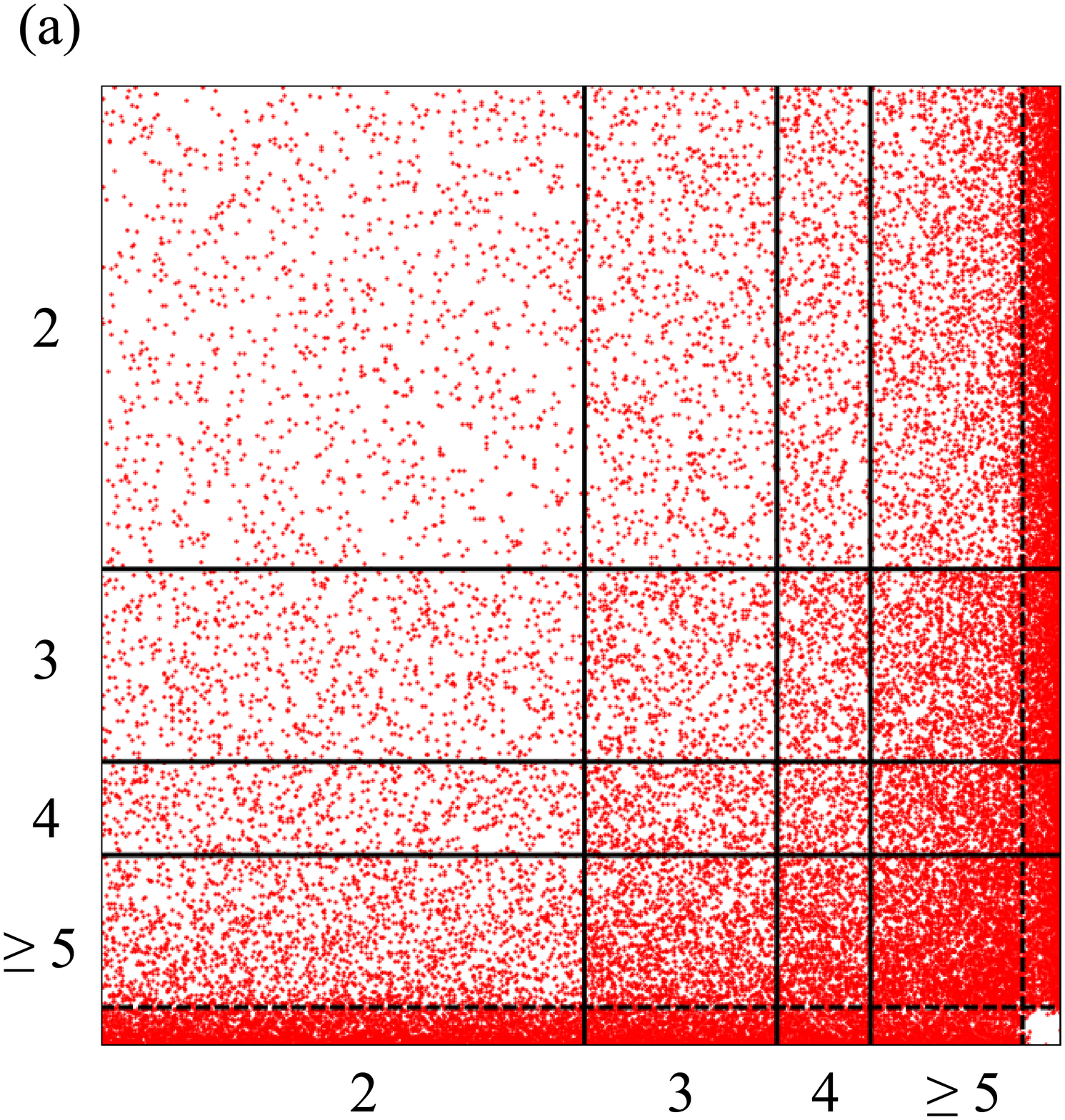}
  \includegraphics[width=75mm]{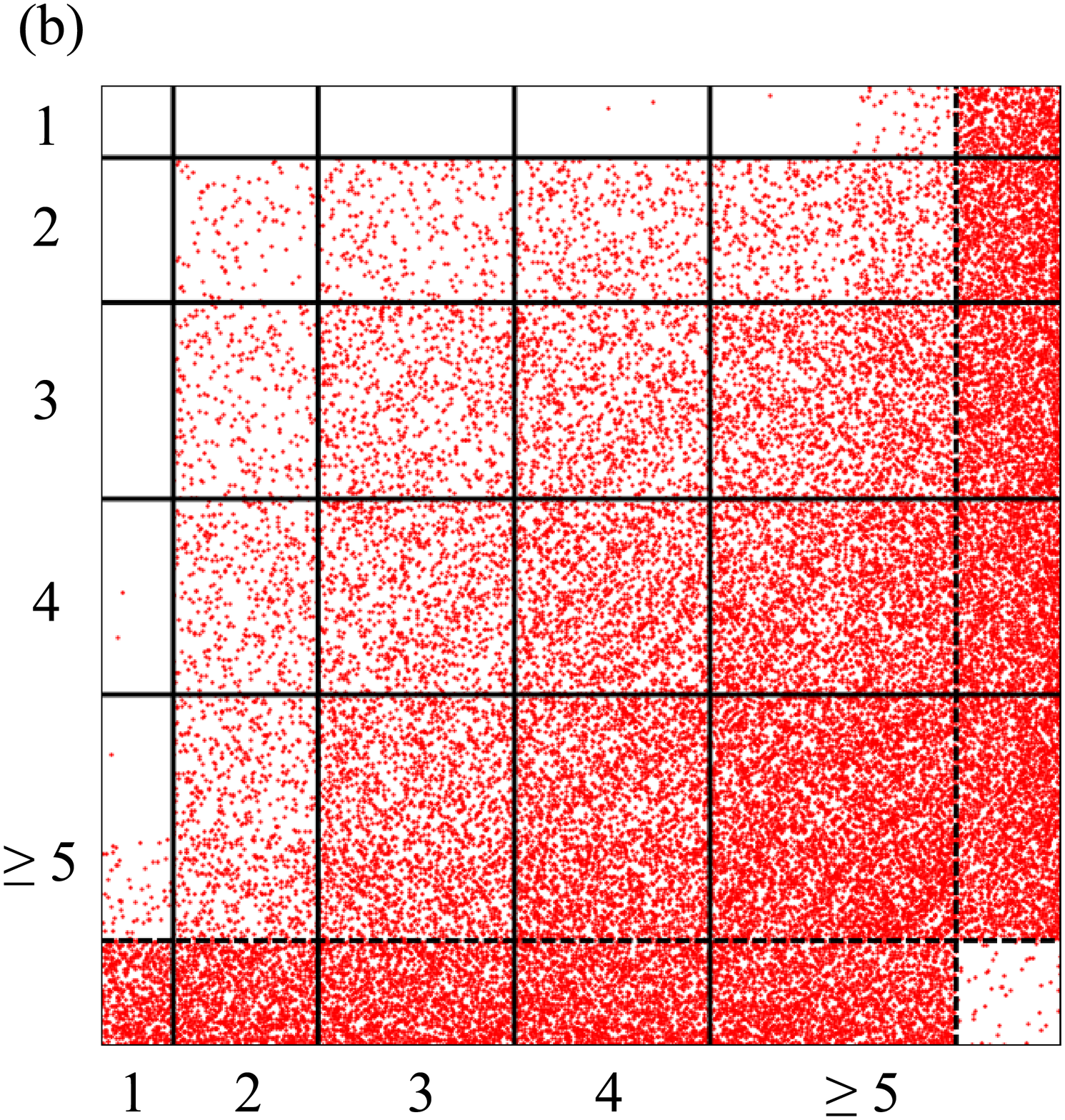}
  \includegraphics[width=75mm]{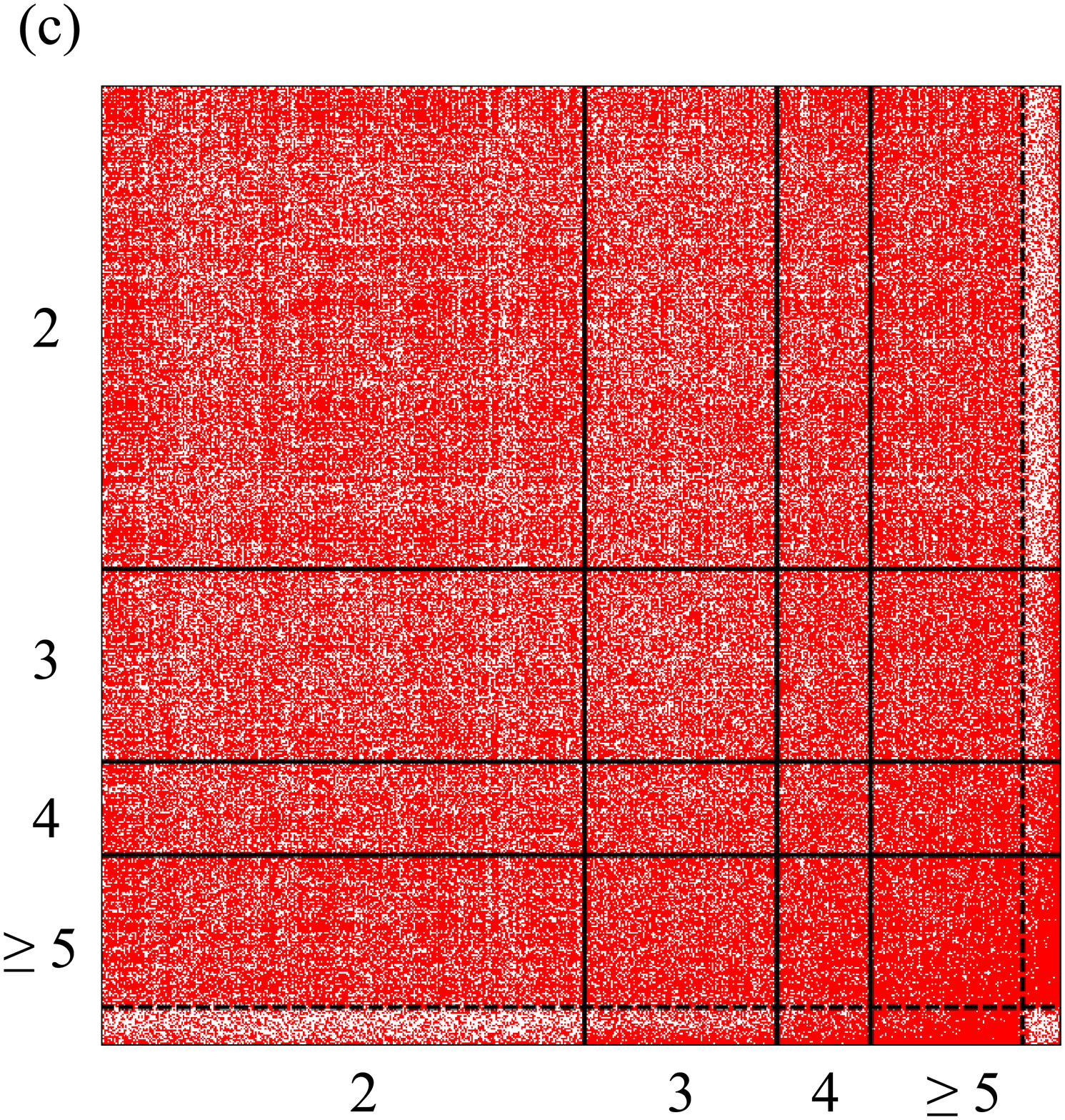}
  \includegraphics[width=75mm]{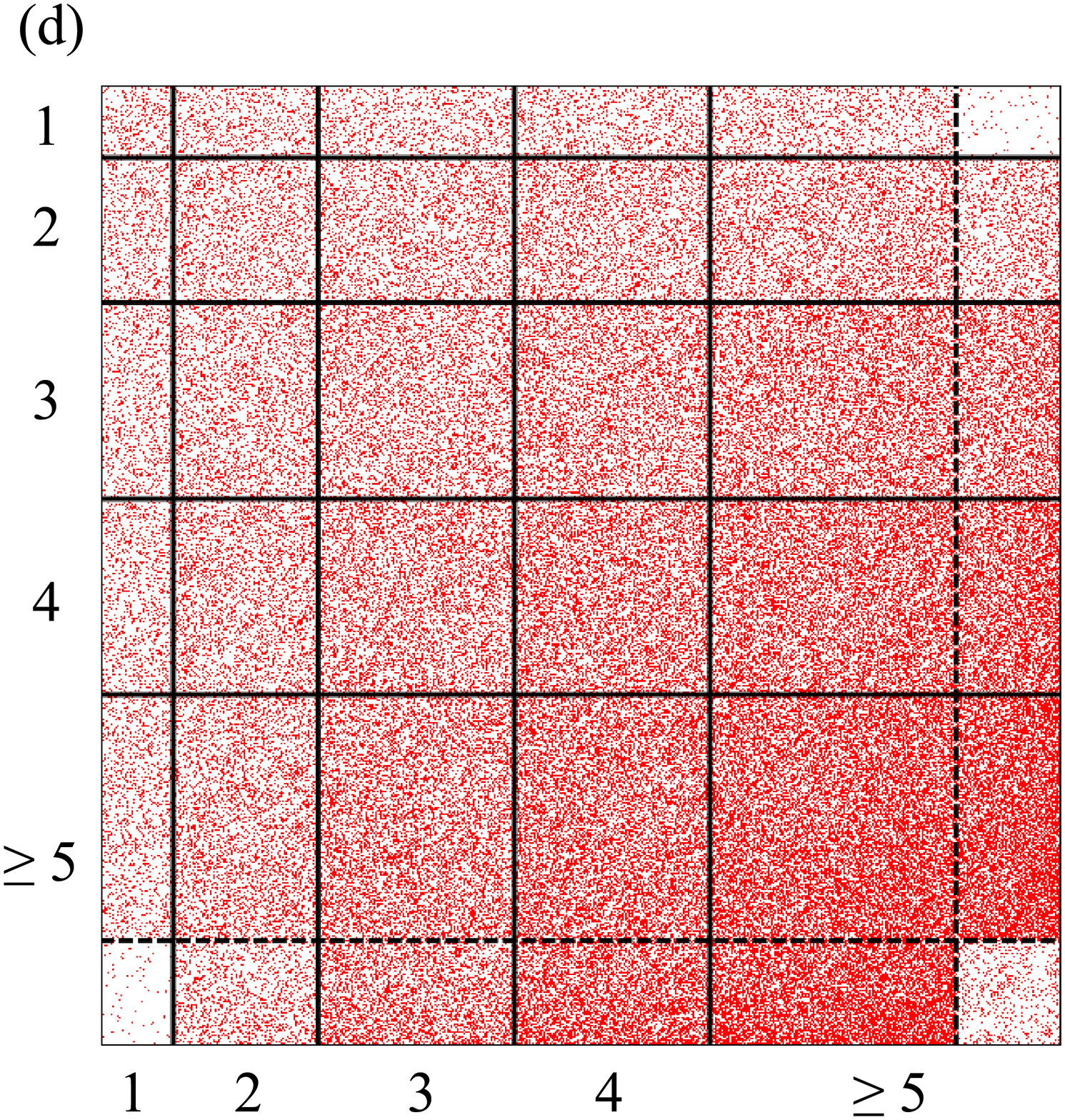}
 \end{center}
 \caption{(Color online) Connectivity matrices after the optimization in terms of $R_{\rm O}$.
(a) Adjacency matrix
for an optimized scale-free network with $p(k)\propto k^{-3}$.
A dot indicates that the corresponding node pair is adjacent.
The nodes are arranged in the ascending order of degree. 
The blocks separated by solid lines represent different degree classes.
The nodes in the same block have the same degree except that we merged the nodes with large degrees into one block. The dotted lines represent node index $i_0$
determined by
$\sum_{i=i_0}^N(k_i+1) \approx N$.
(b) Adjacency matrix
for an optimized Poisson network.
The nodes with $k=0$ are not shown.
For both networks, we set $N=10^4$. 
(c) Connectivity matrix with distance two for the optimized scale-free network shown in (a). A dot indicates that the corresponding node pair is connected with distance two.
(d) Connectivity matrix with distance two for the optimized Poisson network shown in (b).}
 \label{fig:Aij}
\end{figure}

Finally, the adjacency matrices obtained after the optimization in
terms of $R_{\rm O}$ are shown in
Figs.~\ref{fig:Aij}(a) and \ref{fig:Aij}(b)
 for a scale-free network with $p(k)\propto k^{-3}$
and a Poisson network, respectively. In each panel, the nodes are
arranged in the ascending order of degree. The vertical and
horizontal solid lines separate the degree groups such that the
nodes that fall in the same partition have the same degree. This
grouping is done only for small values of $k$ for clarity. Figures~\ref{fig:Aij}(a) and \ref{fig:Aij}(b) indicate
that the optimized networks have little direct connection between
a certain number of the largest hubs. This is the main reason for the negative degree
correlation observed in Fig.~\ref{fig:optimize vary gamma}.

We do not apparently see much structure in the adjacency matrices
shown in Figs.~\ref{fig:Aij}(a) and \ref{fig:Aij}(b) except that hubs
are not adjacent. This result is in contrast with the connectivity
matrices for other negatively correlated networks, such as networks
with maximally negative degree correlation \cite{Menche2010},
fractal networks \cite{Song2006}, and other
hub-repulsive real-world networks \cite{Maslov2002,Maslov2004}.
We consider that the adjacency
matrices shown in Figs.~\ref{fig:Aij}(a) and \ref{fig:Aij}(b) are
produced for the following intuitive reason.  Once a sufficient number
of the largest hubs receive sensors, most of the remaining nodes are
indirectly observed through a hub such that the details of the
connectivity between the remaining nodes little contribute to the size
of the LOC.  With this density of sensors, the neighborhoods of hubs
would be adjacent to each other to make the LOC large. The
number of hubs that are repulsive to each other can be estimated as
follows.  A sensor located at node with degree $k$ makes $k+1$ nodes
observable.  
Because a large $R_{\rm O}$ value implies that the size of the LOC
reaches $\approx N$ at a small $\phi$ value, the
minimum degree of the node with the sensor, denoted by $k_{i_0}$ is
estimated to be the largest $i_0$ such that $\sum_{i=i_0}^N(k_i+1)$
exceeds $N$, where the nodes are arranged in the ascending order of
degree. The $i_0$ values estimated in this manner are shown by the dotted
lines in Fig.~\ref{fig:Aij}. The estimation accurately
describes the number of hubs that are not adjacent to each
other.

This estimation method implicitly assumes that
hubs are separated by distance three. Otherwise, the neighborhoods of different directly observable nodes would overlap even at small $\phi$. Such an overlap
leads to a redundant usage of sensors.
To verify this point,
we plot the so-called connectivity matrix with distance two
for the scale-free and Poisson networks in
Figs.~\ref{fig:Aij}(c) and \ref{fig:Aij}(d), respectively.
By definition, the ($i, j$) entry of this matrix is equal to
unity if nodes $i$ and $j$ are connected with distance two and zero otherwise.
Figures~\ref{fig:Aij}(c) and \ref{fig:Aij}(d) indicate that
the largest hubs are less frequently connected with each other than other node pairs
with distance two as well as node pairs with distance one. This result justifies our estimation method for $i_0$.

\section{Discussion}

We examined observability transitions in correlated networks. We showed that
the LOC is larger for various values of the sensor density (i.e., $\phi$)
for networks with negative degree correlation than those with null or positive degree correlation under both random and degree-based placement protocols. The effect of the degree correlation is more pronounced for degree-based than for random placement. 

In percolation transitions on networks, negative degree correlation tends to shift
the percolation threshold such that emergence of the giant component
requires a larger occupation probability of nodes or links in
negatively correlated networks \cite{Newman2002,Vazquez2003,Noh2007}.
Links between hubs are beneficial in maintaining a
large connected component at least near the percolation threshold.  In contrast, in the model of observability
transitions, negative degree correlation increases the size of the LOC
even in the case of random placement.  Links between hubs are now
detrimental owing to overlap of the neighborhoods of
hubs, which implies redundant usage of the sensors.  If pairs of hubs
are within distance three, the hubs' neighborhoods are easily
connected, resulting in a large LOC.

Networks optimized in terms of the size of the LOC under degree-based
sensor placement have negative degree correlation. In fact, the main
cause of the negative degree correlation is the extremely hub-repulsive
structure of the final networks (Fig.~\ref{fig:Aij}). This structure
is distinct from that of other types of networks with negative degree
correlation.  First, the networks with maximally negative correlation
in terms of $r$ have a bilayer structure~\cite{Menche2010}.  In these
networks, the nodes with the smallest degrees are exclusively connected
with the largest hubs, those with the second smallest degrees are
exclusively connected with the largest hubs among the remaining nodes,
and so on.  Second, fractal networks also have negative degree
correlation~\cite{Yook2005,Song2006,Rozenfeld2007NewJPhys,Guimera2007NatPhys,Palotai2008IUBMB}.
However, the so-called correlation profile, i.e., an averaged
adjacency matrix in which the nodes are ordered in the ascending order
of degree, for fractal networks \cite{Song2006} is not similar to
the adjacency matrices for our optimized networks
(Figs.~\ref{fig:Aij}(a) and \ref{fig:Aij}(b)). The protein
interaction and gene regulatory networks in yeast
\cite{Maslov2002} and the Internet \cite{Maslov2004} also have
the hub-repulsive property, but their correlation profiles at small
degree nodes are quite different from those of our optimized networks.
The fact that our optimized networks are different from these
negatively correlated networks is consistent with the fact that the
$r$ value alone does not specify the network
structure~\cite{Alderson2007,Serrano2007,Weber2007,Whitney2008,Menche2010,Dorogovtsev2010}.

The rich-club phenomenon is a property of networks in which
hubs are densely interconnected as compared to other pairs of nodes
\cite{Zhou2004,Colizza2006}. In contrast,
links between hubs are extremely underrepresented in our optimized networks. Therefore, our optimized networks could be said to show
an anti-rich-club phenomenon. In fact, the hub-repulsive structure has been suggested to be the inverse of the rich-club phenomenon
\cite{Palotai2008IUBMB}.
Further investigation of this network structure and its relevance to
other phenomena and applications warrants future work.

\begin{acknowledgments}
%
This work is supported by Grants-in-Aid for Scientific Research (No 23681033)
from MEXT, Japan, the Nakajima Foundation, the Aihara Innovative Mathematical
Modelling Project, and the Japan Society for the Promotion of Science
(JSPS) through the Funding Program for World-Leading Innovative R\&D
on Science and Technology (FIRST Program) initiated by the Council
for Science and Technology Policy (CSTP).
TH acknowledges the support through Grant-in-Aid for Young Scientists(B) (No. 24740054) from MEXT, Japan.
\end{acknowledgments}


%
\end{document}